\documentclass{article}
\usepackage[utf8]{inputenc}
\usepackage[margin=1in]{geometry}
\usepackage{url}

\providecommand{\keywords}[1]
{
  \small	
  \textbf{\textit{Keywords:}} #1
}

\title{Barriers and Opportunities to Accessible Social Media \\ Content Authoring}
\author{
        Letícia Seixas Pereira 
        \thanks{lspereira@fc.ul.pt - LASIGE, Faculdade de Ciências, Universidade de Lisboa, Campo Grande, Lisboa, Portugal},
        José Coelho,
        André Rodrigues, \\
        João Guerreiro,
        Tiago Guerreiro,
        Carlos Duarte \\
        \\ LASIGE, Faculdade de Ciências, Universidade de Lisboa 
        \\ Campo Grande, Lisboa, Portugal
        }
\date{}

\begin{document}

\maketitle


\begin{abstract}
User-generated content plays a key role in social networking, allowing a more active participation, socialisation, and collaboration among users.
In particular, media content has been gaining a lot of ground, allowing users to express themselves through different types of formats such as images, GIFs and videos.
The majority of this growing type of online content remains inaccessible to a part of the population, despite available tools to mitigate this source of exclusion.
We sought to understand how people are perceiving these online contents in their networks and how support tools are being used.
To do so, we performed an online survey of 258 social network users and a follow-up interview conducted with 20 of them – 7 of them self-reporting blind and 13 sighted users without a disability.
Results show how the different approaches being employed by major platforms are still not sufficient to properly address this issue.
Our findings reveal that mainstream users are not aware of the possibility and the benefits of adopting accessible practices.
From the general perspectives of end-users experiencing accessible practices, concerning barriers encountered, and motivational factors, we also discuss further approaches to create more user engagement and awareness.
\end{abstract}


\section{Introduction}
Social networks have permeated every facet of modern society daily life. With the recent events of the COVID-19 pandemic, their usage is at record levels \cite{Times2020a}. 
Facebook has reported an increase of approximately 11\% over the previous year reaching 1.73 billion users in the first quarter of 2020 \cite{Facebook2020b}. 
The possibility to engage with one another, while physically distant, is at the moment more relevant than it ever was.
For people with disabilities, these platforms also play an important role in disability advocacy, as it provides a vehicle for meeting new contacts with disabilities, learning about issues and news related to it, and discussing accessibility challenges and solutions for improving social media inclusion \cite{Gleason2020, Voykinska2016, Whitney2019}.

Despite the contributions and improvements promoted by social networks in recent years \cite{WebAIM2017}, main platforms still present substantial accessibility barriers for users with disabilities.
The complexity of their interfaces compared to many typical websites comes from the fact that they are primarily composed of user-generated content. 
As such, for these platforms to be truly accessible they have to go beyond ensuring the content they control and produce is accessible: they need to ensure the content their users produce is also accessible. This is especially relevant for people with visual impairments given the prevalence of user-generated content that is mostly visual (e.g., images, GIFs, videos).
As observed by Voykinska et al. \cite{Voykinska2016}, in order to fully engage with visual content, blind users need to overcome several challenges, in particular, the frequent lack of alternative descriptions in photos, essential to provide them proper contextual information. 
Most of them rely on workarounds such as searching for meta-data (author, geo-localization and even comments posted by other users), or reaching out to a nearby friend or family member to assist them. Conversely, in a survey conducted by Mathur et al. \cite{Mathur2018}, friends and family members of visually impaired users conveyed that writing alternative text is time consuming and requires more thought than inaccessible uploading practices. 

Despite the efforts conducted to improve accessibility of visual content by social network services themselves \cite{Facebook2020c, TwitterAccessibility2020}, little is known about the barriers and motivations to the creation of accessible content by end-users, which is at the root of the problem.
In this research, we sought to explore the current context on the accessibility of visual content in social networks. We provide further analysis on the factors hindering the creation of accessible content by people with and without disabilities on major social media platforms. Furthermore, we also aimed to uncover what does or can motivate people to create accessible content. We set the following research questions:

\begin{itemize}
    \item RQ1: How aware are social network users of content accessibility needs and tools and which barriers they encounter when sharing and authoring accessible media content? 
    \item RQ2: What are the requirements for social network users to create accessible media content?
    \item RQ3: What are the motivations for social network users to create accessible media content?
\end{itemize}

To answer these questions, we conducted a user study with three phases: 1) preliminary interviews; 2) online questionnaires; 3) final interviews. 
First, we conducted preliminary interviews with five participants from a variety of backgrounds, with and without disabilities. 
We relied on these preliminary interviews to identify key points to explore in the next two phases. 
Next, we conducted an online survey on the use, awareness and barriers of media content authoring on social networks gathering a total of 258 valid answers by people with diverse backgrounds and accessibility awareness. 
We followed up with 20 of these participants, 13 non-disabled users and 7 blind users.
Half of them were not used to author accessible content and, for this group, we asked to do so prior to the final interview and take notes about the experience to share later. 
We focused the interview on the participants’ experiences, seeking to identify barriers and underlying motivations for accessible content authoring. 
In the instances where there weren't intrinsic motivations to pursue accessibility, the interview shifted to understand what could be done, and how to prompt compliance.

Our findings suggest that users are interested in providing accessible content on their social networks but most of them are not aware of the approaches needed in order to improve their practices. 
Current features provided by major platforms are not providing proper support for media consumption by blind users nor properly assisting users into accessible practices. 
Our work leads to a better understanding of the current state of media accessibility on major social platforms.
It also highlights gaps and opportunities to enable a better interaction for blind people and more engagement in accessible practices by other end-users.

\section{Related work}
Our research is related to prior work on (1) existing accessibility approaches employed by major platforms and its impact on visually impaired users' media interpretation, (2) practices on accessible content sharing by end-users, and (3) current advances and gaps on image captioning.

\subsection{Accessibility on social networks}
Social networking services, such as Facebook, have a high adoption rate among blind users \cite{Wu2014}.
This also happens with Twitter, which evolved from a very simple and text-based interface to one that is now filled with multimedia content \cite{Brady2013a}.
The widespread usage of mobile phones contributes to the growth in publications containing visual content.
However, this user-generated content pushes social networks to become increasingly inaccessible \cite{Brady2013a}.

Numerous efforts have been made to improve accessibility of visual content on social networks with some of these initiatives coming from the service providers themselves. 
For instance, in 2016, Twitter included a feature allowing users to compose their own alternative descriptions for their images. 
However, users claim that this initial feature had its drawbacks as it had to be enabled by the users themselves, was hard to find, and to understand \cite{Twitter2016}. 
Even though this changed in 2020 \cite{TwitterAccessibility2020} --  at the moment this resource is active and available by default -- the impact of this measure has not yet been assessed nor discussed. 
Meanwhile, Facebook made a choice to use automatic descriptions by tagging each image uploaded using image detection and recognition algorithms and enabling the user to edit the automatically provided alternative description \cite{Facebook2020c}.

Despite these efforts, as described in the most recent studies on this context, blind and visually impaired users still encounter significant barriers in interpreting these contents \cite{Brinkley2017, Gleason, MacLeod2017, Morris2016, Whitney2019}.

\subsection{Content sharing}

One of the greatest challenges today is the engagement of users in providing accessible content. Through an analysis of over a million images posted on Twitter, Gleason et al.~\cite{Gleason} observed that only 0.1\% contained an alternative text. 
Twitter itself may be accountable for this low number by not enabling by default the feature that allows the inclusion of the alternative description for many years. However, the authors also observed that even the users who enabled this feature to provide the alternative text descriptions did not always write them.

Visually impaired users engage in major photo-related activities as actively as other users, considered by them as part of the social network experience. 
However, practices such as taking and editing photos, and providing an alternative text, often involve undertaking workarounds or getting help from trusted sighted people \cite{Mathur2018, Voykinska2016, Wu2014, Bennett2018}. 

Mathur et al. \cite{Mathur2018} observed that friends and family members of visually impaired users engage frequently in accessible practices.
Also, according to Wu et al. \cite{Wu2014}, users with visual impairments are much more likely to have friends who are also visually impaired.
These two factors seem to play a role in the increased accessibility of images -- in particular, containing or not alternative descriptions -- that are accessed by visually impaired people (in comparison to those that are not) \cite{Mathur2018}.
However, users also report that writing alternative text is time consuming and requires more thought than inaccessible uploading practices, which explains -- at least in part -- the low percentage of images with an alternative description \cite{Gleason}, previously mentioned.

Concerning users’ engagement, previous works suggest that users who currently provide alternative descriptions on their images are mainly motivated by personal connections to someone with a disability or by a general matter of inclusion \cite{Gleason, Sacramento2020}. 
The authors also reinforce the need of educating users in providing alternative descriptions as well as investigate further approaches on using automated description techniques to improve descriptions provided by authors.
While these findings provide important insights into authoring and sharing practices for accessible content, they leave some unanswered questions around the perspective of end users who are not yet aware of current accessibility approaches and the specific needs of users with disabilities. 
In this paper, we focus on further exploring the challenges these users encounter when trying to engage in such activities for the first time and how to motivate them to create accessible media content.

\subsection{Image captioning}

The low user compliance in providing alternative text descriptions is a common web accessibility problem and therefore some alternative methods are usually employed to fill this gap. 
Stangl et al. \cite{Stangl2020} classified some of the existing approaches to generate image descriptions as human-powered approaches, automated image descriptions approaches and hybrid image description technologies. 
Human-powered are recognised by users for their accuracy and quality of responses. 
Techniques such as Crowdsourcing may have slow response times for real-time needs and a high financial cost. 
Friendsourcing may improve the quality and trustworthiness of the answers received, as friends would better understand the question asked, and also removing financial costs of the service.
Although, the social costs of exposing one’s problems and vulnerability are a serious concern for these users as they may appear or feel less independent \cite{Brady2013a}.
As for automated approaches, unlike the previous technique, they are fast and cheap, allowing platforms to deploy them at scale \cite{Duarte2019, Low2019a}. 
While there are significant efforts undertaken over the past years on image understanding and automated captioning, the accuracy of these captions may not be yet sufficient. 
Caption and phrasing models have an important impact in scepticism as blind and visually impaired people are likely to rely more on automatically generated captions than on their intuition, making decisions based on misinformation~\cite{Salisbury2017,MacLeod2017}. 
Besides that, limitations still exist on the adequacy of these systems used in the open world, especially when captioning the wide variety of images posted to social media \cite{Salisbury2017}. 
In order to fill the gap in image captions, hybrid image description technologies propose a combination of automatic techniques and human intervention to explore a trade-off between both techniques, as explored in \cite{Low2019a, Guinness2018, Morris2018, Salisbury2017}.

Despite prior efforts in evolving alternative interpretations for visual content, user-generated content still has a great impact on the accessibility of media content in social networks. 
For this research, we collected feedback from users about their difficulties and possible motivations in accessible authoring practices.
From that, we provide insights into how this interaction flow can be improved to support their needs, as well as to better engage them in the authoring of accessible media content.

\section{Method}

In order to address our research questions, the study was structured into three different phases. Ethical approval to run the study was granted by our university's Ethics committee.

\subsection{Procedure}

\paragraph{Preliminary interviews to inform survey design:}
We conducted an initial interview with 7 social network users, including three blind people, with a variety of occupations and accessibility awareness. 
Interviews were remotely conducted, taking between 30 and 40 minutes. 
A preliminary version of the online questionnaire was applied.
Questions were built around the main topics of this research, such as their usage of social networks, motivations and barriers when authoring accessible media content, and awareness of accessible practices.
Participants were encouraged to propose suggestions and improvements to enhance the overall understanding of the survey. 
Based on their contributions and insights, some adjustments were made in order to construct the following survey and final interviews.

\paragraph{Online questionnaires:}
We conducted an online survey to gather data as well as to recruit participants to the subsequent phase. The specific questions can be found in the Appendix.
The survey was built using Microsoft Forms in four different languages (English, French, Portuguese and Spanish) in order to reach a diverse sample of participants. 
The survey took about 15 minutes to complete and featured a brief section regarding participants' demographics and technology use, followed by some short open questions about authoring practices in social networks. 
Participants were invited to share some additional thoughts as well as to share their emails before finishing the questionnaire, if interested in participating in the next phase. 
We disseminated the research through the research team social media contacts, including their institutions and fellow organisations (including disability-related ones). 
We gathered a total of 258 answers from participants aged from 17 to 73 years old (Mean=37.35, Median=31, IQR=23) with 64 (25\%) of them self-reporting having some kind of disability, such as visual, hearing, motor and/or cognitive impairments.
In particular, 34 participants were blind, while 12 had low vision.
Even though this research is focused on blind people, accessibility in media content also has an impact on the interaction of people with other types of disabilities, for instance, video captions for users with hearing impairments \cite{Berke} or alternative descriptions for other screen reader users, such as people with low vision \cite{Bennett2018} or cognitive impairments \cite{Hemsley2015, Hynan2015}.

\paragraph{Final interviews:}
We followed up with 20 survey participants in semi-structured interviews.
From these participants, 7 self-reported being blind and 13 of them were sighted users without a disability, as summarised in Table \ref{tab:demoParticipants}. 
Moreover, half of interviewees stated they were not normally involved in accessible practices.
As such, they were asked to post some media content on their usual social networks in an accessible way. 
In addition, they were also asked to take notes reporting in detail the activities conducted and their opinions and difficulties encountered in this process. 
We asked questions about their experience in accessible practices in social networks, motivations for accessible content authoring and potential suggestions or additional thoughts on how to improve this process and to more fully commit end-users to accessibility practices.
All interviews were conducted remotely over the phone, Skype, or Zoom and lasted 20 to 30 minutes.

\begin{table}[ht]
\centering
\caption{Demographics of interviewees including age, most accessed social networks, social networks users most post their content, and social networks users most share their content.}
\label{tab:demoParticipants}
\begin{tabular}{|l|c|c|c|c|c|}
\hline
ID   & 
\multicolumn{1}{l|}{Age} & 
\multicolumn{1}{l|}{Accessible practices} & 
\begin{tabular}[c]{@{}c@{}}Most accessed \\ social network\end{tabular} &
\begin{tabular}[c]{@{}c@{}}Social networks users \\ most post their content\end{tabular} &
\begin{tabular}[c]{@{}c@{}}Social networks users \\ most share their content\end{tabular} \\ \hline

\multicolumn{6}{|l|}{Blind participants} \\ \hline
BP1 &   20  &   Yes &   Facebook    &   Facebook    &   Facebook    \\ \hline
BP2 &   63  &   Yes &   Facebook    &   Facebook    &   Facebook    \\ \hline
BP3 &   53  &   Yes &   Twitter     &   Facebook    &   Facebook    \\ \hline
BP4 &   52  &   Yes &   Twitter     &   Twitter     &   Twitter     \\ \hline
BP5 &   50  &   Yes &   Twitter     &   Twitter     &   Twitter     \\ \hline
BP6 &   21  &   No  &   WhatsApp    &   WhatsApp    &   WhatsApp    \\ \hline
BP7 &   17  &   Yes &   Twitter     &   Messenger   &   Twitter     \\ \hline

\multicolumn{6}{|l|}{Sighted participants}  \\ \hline
SP1     &   73  &   Yes &   Twitter     &   Twitter     &   Twitter     \\ \hline
SP2     &   32  &   No  &   WhatsApp    &   WhatsApp    &   Instagram   \\ \hline
SP3     &   30  &   No  &   Facebook    &   Facebook    &   Facebook    \\ \hline
SP4     &   57  &   Yes &   Twitter     &   Facebook    &   Twitter     \\ \hline
SP5     &   30  &   No  &   WhatsApp    &   Instagram   &   Instagram   \\ \hline
SP6     &   25  &   No  &   Instagram   &   Instagram   &   Instagram   \\ \hline
SP7     &   30  &   No  &   Twitter     &   Messenger   &   Messenger   \\ \hline
SP8     &   33  &   No  &   Facebook    &   Facebook    &   Facebook    \\ \hline
SP9     &   41  &   No  &   Facebook    &   Facebook    &   -           \\ \hline
SP10    &   27  &   Yes &   Instagram   &   Instagram   &   Instagram   \\ \hline
SP11    &   29  &   No  &   WhatsApp    &   WhatsApp    &   Instagram   \\ \hline
SP12    &   34  &   Yes &   Instagram   &   Instagram   &   Instagram   \\ \hline
SP13    &   30  &   No  &   WhatsApp    &   WhatsApp    &   WhatsApp    \\ \hline
\end{tabular}
\end{table}

\subsection{Data analysis}
In order to analyse the gathered data, all conducted interviews were transcribed and two researchers reviewed independently a subset of them through a mixed inductive and deductive coding approach.
The coders compared their initial set of codes and their categorisations in order to develop a unified list of codes.
Then, the two coders reviewed a new subset of transcriptions taking into account the consolidated and revised codes in order to reach an agreement, identifying a total of 150 distinct codes.
Cohen’s Kappa (K) \cite{Cohen1960} was calculated on agreement between the two coders and results indicated that a good level of agreement was reached (K=0.70).
Following that, the coding of all data collected was performed, including all of the survey's open questions.

\section{Findings}

In this section, we reflect on the relevant themes that emerged during our analysis of the data gathered from the survey and the interviews conducted. 
We divided our findings as follows: 
(1) social media usage, 
(2) unaware of accessibility, 
(3) lack of know-how, 
(4) the cost of the additional effort,
(5) complying with and without guidelines or features, 
(6) inaccessibility, 
(7) and accessibility motivations and concerns.

\subsection{Social media usage}
Concerning the social networking platforms used by our survey respondents, Twitter was the first choice of the majority of users with disabilities, followed by Facebook.
As for the respondents without disabilities, Instagram was their first choice, also followed by Facebook.
The popularity of Twitter among users with disabilities may be explained by the simple text-based interface initially provided by the platform \cite{Brady2013a}, gathering many users with disabilities over the years.

Participants with disabilities are creating and sharing more non-visual content, such as text and audio, than participants without disabilities.
As for visual content, participants with disabilities share more videos, while participants without disabilities share more images, as also observed by Wu et al.~\cite{Wu2014}. 
Even though visually impaired users edit and post photos on social media, they still encounter several challenges using these apps \cite{Bennett2018}, which may be contributing to their low participation in image-centric social networks such as Instagram. 
In particular, among the type of media mentioned, GIFs and Memes were highly unpopular among respondents with disabilities, with most of them reporting never sharing this type of content.
As explored in previous works, these kinds of media represent a challenge for accessibility.
Besides not being properly supported by major platforms \cite{Sacramento2020}, they carry some cultural context or hidden meaning related to an emotional tone or humorous aspect \cite{Gleason2019a, Gleason2020a}.
Thus, making it difficult to convey the full meaning of these contents by automatic tools.

Our survey respondents are more frequently engaging in social networks through mobile applications provided by the major social networks. 
Also, visually impaired participants reported using the mobile version of platforms' websites. 
As pointed out in previous works \cite{Voykinska2016, Brinkley2017}, many blind users choose these interfaces for their simplicity, over their desktop counterparts, although mobile versions often have missing features in comparison \cite{guerreiro2013blind,Wentz2011}.

\subsection{Unaware of accessibility}

We found an overall lack of awareness and stigma associated with visually impaired people using technologies, in particular, about accessing visual content.
The vast majority of survey respondents with a disability reported providing an alternative description for their last shared media content while most of non-disabled participants declared not having provided it.
Most of them declared not knowing that it was possible, when asked about the reasons for not engaging in such activity.
Further investigation conducted through interviews allowed us to observe that most of them were not aware of the use of technologies by people with disabilities and how this content is being consumed by them, therefore becoming more difficult to understand what accessible content really means:
\begin{quote}
    “I am describing this image, theoretically for a blind person, but then how will the platform use it? Does the person have the possibility to listen? Is there going to be a...? I don't know how this is used in the end, it seems silly, but how does it reach the person? How is it accessible in fact, at the end of the experience?” – SP13
\end{quote}

A first layer of this unawareness is reflected by the lack of knowledge on what they should do to improve the accessibility of their content. 
Most sighted participants, although frequent users of social networks, only discovered this possibility during this study:
\begin{quote}
    “I don’t know what is to share something in an accessible way, I at least don’t do it and it doesn't seem that the people I know do it, I think it’s because first of all for not knowing.” – SP5
\end{quote}

Following, some of them also felt that it was not necessary to provide an alternative access for their content as they do not know anyone with a disability, underestimating the reach of authoring and sharing activities:
\begin{quote}
    “Personally, I don’t see why, I don’t see any reason to do it because I don’t have anyone who would need this additional explanation either.” – SP5
\end{quote}

However, while some sighted participants were used to privately sharing some family pictures and may follow this line of argument, this misjudgement about the impacts and needs of accessibility may go further, reinforcing the stigma, by implying that visual content is not for blind users:
\begin{quote}
    “I also share a bit of this stigma, which is that, if Instagram is highly visual, a blind person will use it? [...] I don’t know if there is this motivation on the part of the blind people to use a tool that is so visual.” – SP5
\end{quote}

Under the assumption that blind people are not interested or motivated in accessing visual content, the accessibility of this content is not taken into account. 
This is particularly evident in image-centric social networks such as Instagram.
Consequently, blind people are excluded from accessing this content and thus from using these platforms:
\begin{quote}
    “Instagram, well, it's not very accessible for me.” - BP5
\end{quote}

We also observed through survey responses a correlation between adopting accessible practices and the familiarity with the needs of people with disabilities. 
The vast majority of non-disabled participants had no idea about accessibility practices adopted by people around them while half of participants with disabilities stated having friends or family members posting or sharing accessible content. 
Still, 27\% of respondents with a disability convey that people in general do not think that it has a real impact. 
From the perspective of two blind interviewees, some people just don’t care enough about the subject. 
Partly because many of them don't know someone with a disability, so they do not reflect about it, but also because some may be focused on just reaching out to as many people as possible, neglecting minority user groups:
\begin{quote}
    “Other people don’t care, no matter how much we tell them.” – BP5
\end{quote}

\subsection{Lack of know-how}
From those who provided alternative descriptions, most of them used the features provided by the platform itself. 
The rest of them, all sighted interviewees, embed it in the content of their post. These users evoked having discovered alternative descriptions in general through other users' publications and therefore used it as a model example to complete this task, following previous studies \cite{Gleason}. 
This practice, although not ideal, may increase the visibility of issues related to content accessibility. 
This choice was mainly driven by interviewees not perceiving nor discovering the specific accessibility feature provided by the platform. 
Followed by the 21\% of survey respondents that stated not knowing \textbf{where to write} an alternative text description, it reinforces some statements on the not user-friendly and hard-to-find accessibility approaches currently employed by social network service providers:
\begin{quote}
    “If they told me ‘Oh yes, you can make these things accessible, you just have to take a lot of very complicated steps to get it done’. Obviously, this will make it very difficult for people who are interested in doing it. Accessibility features are difficult to find and not very discoverable, they’re hidden.”- SP8
\end{quote}

Furthermore, 14\% of survey respondents also mentioned not knowing \textbf{how to write} a suitable alternative description as a reason for not providing it. 
While interviewees convey that it was a challenge to represent an image through text, they also stated finding no guidance or information on what is considered a good alternative description:
\begin{quote}
    “My biggest difficulty was this, not knowing if there was any standard or not, what would be necessary to include or not, if I should put more information, less information, if I should be more specific, give more details…” – SP13
\end{quote}

From that, it was a consensus among our interviewees that, as a first step, it is essential to create approaches to raise awareness on accessible practices and its benefits between social network users. Most of them suggest that one way to achieve equity in social media is by emphasising accessibility features as, currently, they do not draw enough attention of these users to this matter:
\begin{quote}
    “They have to be more visible in the authoring process […], even if they weren't hidden, they could also be promoted, which they aren't either.” – SP7
\end{quote}

Besides, accessibility must be a part of the authoring process. Interviewees consider that an edit field for the alternative description should be provided within the authoring flow with a content warning not only just to make it mandatory but, in particular, to get people used to providing this information:
\begin{quote}
    “There is a pretty obvious place to put a caption, so why not just have the alt text to be there? If there is a caption field, and there is an alt text field, I think people […] would be ‘oh yeah, I should put an alt text, so you are here to remind me’ […]. Just making the alt text field more up front and maybe even if not required, at least prompted, if you don't enter it, it asks the question, are you sure you don't want to include alt text? Maybe that's a good way to go.” – SP4
\end{quote}

Furthermore, 30\% of interviewees also convey that having a support feature to suggest alternative descriptions would help them, first by providing an example of what is considered an appropriate alternative description, and also by providing the opportunity for them to improve machine generated descriptions:
\begin{quote}
    “This could become standard [...] show the description and say: "Do you think this is a good description for this image?" because in this case, it [Facebook] simply decides the description itself and also, if it is not correct, I have no way of indicating it.” – SP8
\end{quote}

\subsection{The cost of the additional effort}
Mentioned by 15\% of survey respondents, the third most popular reason for not providing an alternative description for their media content was the time it takes to do it. 
Some sighted interviewees also declared that, as accessible practices are currently not integrated into their current publishing interaction flow, providing an alternative description for their images adds a new task that would involve considerable time for reflection and make their activities much more time-consuming, especially given the spontaneous nature of social media content sharing for personal usage:

\begin{quote}
    “Depending on the elements that are in the picture can be a bit exhausting, especially if you have many people and I have to be describing each person …” – SP5
\end{quote}

People with disabilities share the additional burden of often having to be the activist in their social circles, bringing awareness and recalling others to ensure they start producing accessible content.
This leads some to the feeling of being obnoxious for having to constantly remind their acquaintances they are not able to fully understand the content shared:
\begin{quote}
    “When we have a disability and we're faced with it, we're tired of saying ‘hey, don't forget, publish something accessible’.” – BP5
\end{quote}

As reported by our blind interviewees and more broadly discussed in previous works \cite{Voykinska2016, MacLeod2017, Zhao2017, Morris2018, Whitney2019, Stangl2020}, alternative descriptions currently provided by major platforms are not providing enough contextual information for visually impaired users to properly interpret media content.
Therefore, it is essential to promote user engagement to convey this additional context and the users' own intentions and purposes in publishing this content through human description:
\begin{quote}
    “When it comes to being accessible, you can’t take it out of somebody’s hands […] Let them decide what they want to say, [..] how they want to describe it, whatever they were thinking of, why they were posting, that image should be what they’re posting.” – BP4
\end{quote}

\subsection{Lack of standardisation}

Major platforms are employing different approaches to support accessible media authoring. This lack of standardisation requires users to identify and learn how to use each one of the accessible features provided by every platform used.
In Facebook, only two participants publishing accessible content for the first time were able to discover this feature, and, as stated by one of them, after a thorough research:
\begin{quote}
    “On Facebook, that then I really, really had to go search, I tried a lot, I tried and eventually I had to search what Facebook had for accessibility because I could not find it.” – SP7
\end{quote}

They also reported not finding this feature available in the native app, only being able to edit the automatically generated description on their desktop through the web interface.
One participant also shared that, even though he is blind and a frequent Facebook user, he was not aware of the possibility of editing these descriptions.
As for Twitter, the feature enabling the user to provide an alternative description for their own images was considered the best approach by one blind user.
However, as consent by other participants, its effective use is jeopardised by not being part of the standard Twitter authoring process:
\begin{quote}
    “Twitter's way of making a field for people to put in the alt text is the best way to do it, but I can't tell you how many people can find it.” – BP4
\end{quote}

Only one participant tried to explore the accessibility feature provided by Instagram during this study and also mentioned the difficulty in finding it:
\begin{quote}
    “In Instagram, the first thing I think I had heard about is the alternative text, but I tried to figure it out and I found it out, but with some difficulty, because it was in some advanced settings in the post, it's not very visible.” – SP7
\end{quote}

During the interviews, Facebook, Twitter and Instagram were the most cited social networks among participants. 
Each of these platforms adopted a different approach for providing alternative descriptions for user-generated content.
Twitter provides an input field for users, while Facebook and Instagram also rely on machine-generated descriptions.
However, the non-user-friendly interface is a common ground between them.
Therefore, concerning the roles of different stakeholders involved in the accessibility context, a majority of interviewees consider that platforms have the largest influence and, currently, they are not fully engaged in promoting accessibility:
\begin{quote}
    “The platform is responsible for ensuring a good user experience and they have a good percentage of users who have these specific needs, and they are the ones who have to ensure this experience also applies to them.” – SP7
\end{quote}

30\% of interviewees also consider this responsibility should be shared with the users themselves.
As they play an essential role in this process, they have to be more committed with accessible practices.
Finally, 25\% of them consider this matter also involves governments' actions. 
Three participants mentioned the legal context of digital media regulation, and two of them highlighted accessibility legislation for physical structures, such as ramps and handrails.
Two participants also believe it involves providing better public policies for social inclusion in general.
One participant summarised her thoughts on how each one of these three main stakeholders could be contributing to promote accessibility:
\begin{quote}
    “I think that platforms, besides enabling this content, should talk more about this, about accessibility, this being more explicit. I think the actors should be more conscious in this sense. I'm also within this process, I'm questioning the social networks, but I'm not actively participating in this process either. So, I think it's up to an active participation, especially of people with greater reach, of companies and institutions. And I think that, in this sense, governments also have an important role in regulating specific institutions that produce accessible content. In order for us to have this accessibility in a physical space, and, now more than ever, in a virtual space.” – SP13
\end{quote}

\subsection{Inaccessibility}

The inaccessibility of accessibility features is a paradox currently encountered in these services. Just like sighted participants, blind interviewees also reported having difficulties identifying the proper feature provided by major platforms even though they were familiar with its availability. This difficulty is further reinforced by the constant updates of these systems, requiring the acquisition of new knowledge about the structure of these new interfaces, as also pointed out by Voykinska et al. \cite{Voykinska2016}:
\begin{quote}
    “I find it difficult to find the field to put the alt text in, because every time I go into Twitter and send a picture, it’s changed.” – BP4
\end{quote}

As observed by Sacramento et al. \cite{Sacramento2020}, major social network services are not consistent when it comes to providing accessibility features and compliance among different platforms, such as mobile/desktop interfaces or iOS/Android mobile applications.

Furthermore, while most of blind survey respondents declared creating accessible content on their social networks, in the interviews it was possible to observe that this practice is not necessarily being enabled by accessible features provided by platforms. Most interviewees reported asking for help in order to confirm the elements contained in the media to be published, making them depend once again on sighted friends or family members. Moreover, one of them stated only sharing content already accessible. Therefore, these users declared missing a feature to assist them in creating their own descriptions for their images:
\begin{quote}
    “In this case [for this study], I had to ask my brother how the photos were like for me to describe them.” – BP6
\end{quote}

This lack of support for blind authors reinforces the social cost also observed by \cite{Brady2013a} and their sense of exclusion as they are unable to fully experience this other aspect of social networks:  
\begin{quote}
    “I don't think it's because we're blind or have any other type of disability that we lose the right to express ourselves, or to send any kind of joke, let's say, those little stickers... when talking, we who are blind, are excluded from doing it.” – BP1
\end{quote}

As mentioned by BP1, the constant arising of new media content such as GIFs, Stickers, Memes, and images with embedded text or screenshots enabled by major platforms pushes these social networks to be less and less accessible for these users, as this diversity often fails to match the accessibility features available, which allows us to better understand the numbers obtained in the survey:
\begin{quote}
    “[…] but GIFs, I can't picture myself, what's the point or what it does. There are more and more little things like that appearing.” – BP5
\end{quote}

\subsection{Accessibility motivations \& concerns}

Participants currently publishing accessible content are mostly driven by having a disability themselves or having acquaintances with disabilities. Two sighted participants not used to share accessible content declared that they would be more engaged in these practices if they had a personal connection to someone with a disability:
\begin{quote}
    “What would motivate me would be to have a follower who I know is blind, which I don't have.” – SP5
\end{quote}

During the interviews some of them also mentioned acquaintances with disabilities, realising how these practices could also benefit them:
\begin{quote}
    “Actually, I just remembered that yesterday, a friend of mine [...] had an accident and is experiencing visual difficulties because of a trauma. And I thought that, in the future, if he needs and if it helps, I can always add an alternative text to the posts so that at least my friends can have access to it.” – SP7
\end{quote}

All sighted participants declaring publishing accessible content agreed the main reason for them to engage in these practices is because it’s the right thing to do. Two of them work in the context of web accessibility and the other two work with media communication:
\begin{quote}
    “What motivates me is really to promote access, my work line is communication, so it doesn't make any sense if I talk about communication and don't think about exposing it in such a way that as many people as possible are included, as many diversities as possible, people with disabilities, minority groups. So what motivates me is the essence, to be able to extend this process […] to all people.” – SP10
\end{quote}

30 \% of interviewees, all of them not used to share accessible content, declared being interested in contributing to the inclusion of people with disabilities and to make information reach as many people as possible. 
Furthermore, most interviewees also convey that providing platform support is an important factor to create more user engagement. 
One blind participant, who first discovered accessibility features during the study, stated that now he is aware of it, he has already included alternative descriptions for all his previously posted pictures and intends to adopt this practice in the future.
Nevertheless, the difficulty encountered may discourage some of them, causing the opposite effect on people who, at first, are more likely to give it a try.
\begin{quote}
    “I think that if I knew how to do it, I could feel more motivated, if I had these instructions.” – SP13
\end{quote}

It was also possible to observe that some interviewees may consider some setbacks on accessibility approaches.
The strategy of embedding an alternative description text in the post’s content previously mentioned, while it may be perceived by some as an example, was mentioned by one participant as downside of accessibility, as it makes a post very long and somehow redundant for those not using a screen reader. 
For that, she considers that this information should be embedded in a way that is only perceived by screen reader users.
Loading and scroll speed was also considered as a setback by one interviewee as accessible practices concerns including additional information to be loaded by apps and websites:
\begin{quote}
    “It occurred to me soon, people could describe just like I did, or record a short audio to be heard by these people, but this will take a lot out of the dynamism that people have gotten used to in Instagram. […] it would be great for them, but I think it represents a setback for non-visually impaired people.” – SP5
\end{quote}

\section{Discussion}
In what follows, we firstly discuss how these findings can be used to answer our research questions followed by further contributions provided by this work.

\subsection{Research questions}

\subsubsection*{RQ1: How aware are social network users of content accessibility needs and tools and which barriers they encounter when sharing and authoring accessible media content?}

First, it is important to highlight that the barriers encountered and reported by participants are strongly related to their familiarity with people with disabilities, with assistive technology and its behaviours on social networks. 
For this reason, it is extremely important to educate people about disability, including how different disabilities affect the way people interact with technology, their challenges, and how can users publish accessible, inclusive digital content. While the society at large should be responsible for seeking and providing such education, social networking sites may leverage their platform to educate users and enable more inclusive sharing practices. Instead, accessible practices and features remain unknown to most users.
Most sighted participants were \textbf{not aware} of the steps they can take to make their content more accessible and they found \textbf{no guidance on major platforms} to assist them in this process.
In addition, even when they search for guidance they report difficulty in learning about or in finding accessibility features.
Given this context, accessibility practices are being perceived by many users as an activity that \textbf{requires a significant additional effort} on their part.

We also identified that some still have a certain \textbf{stigma associated with accessibility}, arguing that accessibility should be employed only when necessary or, even worse, that accessibility compliance may compromise their current experience on social media. 
Platforms not making this a requirement or not providing a proper prompt warning may also be contributing to this first line of thought. 
On the other hand, there is still a \textbf{lack of support for blind users} to create accessible media content as they find no features to assist them in this activity. 
Even though they recognise the latest advances promoted by major platforms, accessibility issues have been present in these services for a long time \cite{buzzi2010facebook, buzzi2011web} and people with disabilities are still highly dependent on other people to more equally participate in social networks, reinforcing the social cost constantly experienced by them.

\subsubsection*{RQ2: What are the requirements for social network users to create accessible media content?}

As a first step, major platforms must provide a more \textbf{user-friendly and accessible interface} in order to \textbf{make accessibility features more noticeable and easier to use}. Another critical issue identified concerns the different approaches adopted by these platforms to provide accessibility features, making it difficult for users to identify and learn how to use each one of these resources. 
For that, we suggest that \textbf{accessible approaches should be more standardised among platforms} in order to create an \textbf{easily recognised pattern} towards an accessible posting and sharing inherent routine. 
Moreover, machine-generated descriptions introduced by some platforms often do not provide blind users with enough information or context, so they properly understand the corresponding media content, as also observed in previous works \cite{Voykinska2016, MacLeod2017, Zhao2017, Morris2018, Whitney2019, Stangl2020}.
From that, one promising avenue is investing in \textbf{hybrid solutions}: exploring the balance of the benefits of technological advances in automatic recognition and machine-generated descriptions, and involving users to fill the gap concerning context details and, in particular, their own intention and purpose in posting a certain media content. 
This approach also harnesses automatic descriptions to guide and support users in creating their own description, reducing the work and time cost perceived by some of them.
Furthermore, blind users authoring accessible content will also benefit from it, reducing the social cost previously discussed.

\subsubsection*{RQ3: What are the motivations for social network users to create accessible media content?}

Our participants were mainly motivated by \textbf{doing the right thing} and \textbf{promoting inclusion} for people with disabilities. 
Although some of them are used to share media content only for a private audience, such as family and close friends, many of them are interested in enabling access to information for other people as well. 

Considering this willingness in including people with different abilities and cultural contexts shown by interviewees, some strategies may be applied in order to motivate more users to create accessible media content.
Participants tended to be more aware when in contact with a person with a disability, but as well when confronted with current accessibility approaches. 
They showed curiosity and interest in knowing more about the subject and they are motivated to understand how their content is being consumed by people with disabilities.
One major challenge identified among interviewees was the unawareness on how and why people with disabilities use social networks.
By \textbf{making users part of this process}, integrating accessibility features more prominently on platforms authoring flow, and \textbf{educating them about accessible practices and alternative access}, providing tutorials or scenarios of people with disabilities using the Web, for instance, it may increase awareness and thus encourage them to become more frequently engaged in such practices.
From that, we reinforce the conclusions reached by previous studies \cite{Gleason, Sacramento2020} on the need of additional tooling and training for social network users on all major platforms.

\subsection{Perspectives on accessible social media content authoring}

The different user perspectives on accessible practices in social networks gathered in our study allowed us to identify potential avenues for future research in accessible media content authoring.

Although machine-generated descriptions are not currently providing sufficient contextual information on visual content, their main advantage is that they can potentially be deployed on a large scale.
For that, research on text alternatives best practices and users preferences \cite{Petrie2005, Salisbury2017, Stangl2020} may be used to further improve these descriptions so that their quality becomes acceptable.
From a hybrid perspective, providing suggestions on appropriate alternative descriptions may be useful to educate users and, therefore, to create more engagement on accessible practices.

Another opportunity suggested by some of our sighted interviewees is providing users with different ways of including alternative descriptions, such as audio descriptions. 
Previous studies \cite{DosSantosMarques2017, Kim2020} identified that human narration -- besides being faster -- allows better image comprehension by blind users and it helps to establish a connection between user and content author.
Marques et al. \cite{DosSantosMarques2017} also suggest a scenario where blind users could send a request to the author of an image so that he would record an audio description. 
While this solution possibly reinforces the existing burden imposed on blind people, collaborative approaches on alternative descriptions are also an interesting aspect to be further explored. 
As suggested by Sacramento et al. \cite{Sacramento2020}, providing users the possibility of including alternative descriptions for visual content they encounter may be employed by sighted users who are already motivated and currently engaged in accessible practices.
This approach could be also a complementary action to create more awareness among social media users.

Moreover, even though this work focused on visual disabilities, most topics discussed in our research questions also apply to different kinds of impairments, in particular concerning accessibility awareness and motivational strategies.

\section{Conclusions}
In this study, we presented an overview of the usage and the accessible practices employed by 258 inquired social network users.
These findings suggest that people with disabilities are interacting in social networks as much as users without disabilities.
However, our participants with disabilities showed being more frequently engaging with non-visual content, such as text and audio, than non-disabled participants.
This may be partly explained by another finding: most participants without disabilities are not sharing accessible media content, because, as reported by most of them, they were not aware of this possibility until participating in this study.

Following that, we conducted interviews with 20 of these respondents in order to better explore their experiences, challenges and motivations concerning accessibility and visual content in social networks.
While non-disabled interviewees were interested in being more engaged in accessible practices, most of them did not find proper assistance or support on major platforms to guide them to enhance the accessibility of their content.
At the same time, blind users are not being provided with proper alternative descriptions whether by authors or by the machine-generated descriptions provided by some major platforms.
Moreover, the burden of educating others and promoting the authoring and sharing of accessible content falls, unfairly, upon their ability to compel others to act.

Our work also complements previous research on visual content in social networks by providing insights on how non-disabled users are experiencing accessible practices, but also brought up the importance of employing hybrid solutions to fill the current gap. 
Platforms must go beyond just deploying accessibility features and must be more invested in approaches that create user awareness, engaging more people in adopting accessible practices in their daily posting routine.

\section*{Acknowledgements}
This paper was written with support from the SONAAR Project, co-funded by the European Commission (EC). H2020 GA LC-01409741. This work was supported by FCT through funding of LASIGE Unit R\&D, Ref. UIDB/00408/2020.

\section*{Disclosure statement}
No potential conflict of interest was reported by the authors.

\bibliographystyle{unsrt}
\bibliography{main}

\section*{Appendix: Survey questions}

\subsection*{Demographic questions}
\begin{enumerate}
    \item Age
    
    \item Gender
    \begin{enumerate}
        \item Female
        \item Male
        \item Prefer not to answer
        \item Other: 
    \end{enumerate}

    \item Which operating system do you use (Choose all that apply)
    \begin{enumerate}
        \item Windows
        \item MacOS
        \item Linux
        \item iOS
        \item Android
        \item Other: 
    \end{enumerate}
    
    \item Do you have some type of disability?
    \begin{enumerate}
        \item Yes
        \item No
        \item Prefer not to answer    
    \end{enumerate}
    
    \item Which ones? (Choose all that apply)
    \begin{enumerate}
        \item Blindness
        \item Low vision or Visual impairment
        \item Cognitive
        \item Deafness or Hard-of-hearing
        \item Motor
        \item Prefer not to answer
        \item Other: 
    \end{enumerate}
    
    \item Do you use any assistive technology?
    \begin{enumerate}
        \item Yes
        \item No
        \item Prefer not to answer
    \end{enumerate}
    
    \item Which ones? (Choose all that apply)
    \begin{enumerate}
        \item Screen reader
        \item Braille display
        \item Screen magnifier
        \item Prefer not to answer
        \item Other: 
    \end{enumerate}
    
    \item Do any of your family members, friends or acquaintances have a disability?
    \begin{enumerate}
        \item Yes
        \item No
        \item Prefer not to answer    
    \end{enumerate}    
    
    \item Which ones? (Choose all that apply)
    \begin{enumerate}
        \item Blindness
        \item Low vision or Visual impairment
        \item Cognitive
        \item Deafness or Hard-of-hearing
        \item Motor
        \item Prefer not to answer
        \item Other: 
    \end{enumerate}
    
    \item Are you aware of any of your contacts on social networks (friends, followers, etc.) having a disability?
    \begin{enumerate}
        \item Yes
        \item No
        \item I don't know
        \item Prefer not to answer    
    \end{enumerate}    
    
    \item Which ones? (Choose all that apply)
    \begin{enumerate}
        \item Blindness
        \item Low vision or Visual impairment
        \item Cognitive
        \item Deafness or Hard-of-hearing
        \item Motor
        \item Prefer not to answer
        \item Other: 
    \end{enumerate}
    
\end{enumerate}

\subsection*{Social networks usage}
For the next questions, we will use the term Social Networks as a reference to online platforms used by people to communicate or to build social relationships with other people. Some examples of these online services are: Facebook, Twitter, Instagram, WhatsApp, WeChat, among others
(\url{https://en.wikipedia.org/wiki/Social\_networking\_service\#Largest\_social\_networking\_services}).
We will also distinguish some of the activities carried out on these social networks:
- Post your content refers to sharing new content of your own authorship. For example, posting an image that you created on your Twitter or Instagram timeline as well as sending a message on WhatsApp or Messenger.
- Share content from others refers to sharing content authored by another person. For example, sharing on your Twitter or Instagram timeline an image posted by someone else as well as forwarding a message in WhatsApp or Messenger.

\subsubsection*{Access}
\begin{enumerate}
  \setcounter{enumi}{11}
  
  \item How frequently do you use a smartphone or tablet to access social networks?
    \begin{enumerate}
        \item Several times per day 
        \item Once a day
        \item Several times per week 
        \item Several times per month 
        \item Less than once a month 
        \item Never
    \end{enumerate}
    \item How frequently do you use a desktop or laptop computer to access social networks?
    \begin{enumerate}
        \item Several times per day 
        \item Once a day
        \item Several times per week 
        \item Several times per month 
        \item Less than once a month 
        \item Never
    \end{enumerate}
\end{enumerate}

\subsubsection*{Posting your content}
\begin{enumerate}
  \setcounter{enumi}{13}
  
  \item How frequently do you use a smartphone or tablet to post your content on social networks?
    \begin{enumerate}
        \item Several times per day 
        \item Once a day
        \item Several times per week 
        \item Several times per month 
        \item Less than once a month 
        \item Never
    \end{enumerate}
    \item How frequently do you use a desktop or laptop computer to post your content on social networks?
    \begin{enumerate}
        \item Several times per day 
        \item Once a day
        \item Several times per week 
        \item Several times per month 
        \item Less than once a month 
        \item Never
    \end{enumerate}
\end{enumerate}

\subsubsection*{Sharing content from others}
\begin{enumerate}
  \setcounter{enumi}{15}
  
  \item How frequently do you use a smartphone or tablet to share content from others on social networks?
    \begin{enumerate}
        \item Several times per day 
        \item Once a day
        \item Several times per week 
        \item Several times per month 
        \item Less than once a month 
        \item Never
    \end{enumerate}
    \item How frequently do you use a smartphone or tablet to share content from others on social networks?
    \begin{enumerate}
        \item Several times per day 
        \item Once a day
        \item Several times per week 
        \item Several times per month 
        \item Less than once a month 
        \item Never
    \end{enumerate}
\end{enumerate}

\subsubsection*{Access}
\begin{enumerate}
  \setcounter{enumi}{17}
  
  \item How do you access social networks? (Choose all that apply)
    \begin{enumerate}
        \item Website on mobile device or tablet 
        \item Website on desktop or laptop computer 
        \item Mobile app
        \item Other:
    \end{enumerate}

\end{enumerate}

What social networks do you access and how often do you access it? (Provide information about the 3 social networks you access the most)
\begin{enumerate}
  \setcounter{enumi}{18}
    \item First most accessed social network:
    \item Frequency of access:
    \begin{enumerate}
        \item Several times per day 
        \item Once a day
        \item Several times per week 
        \item Several times per month 
        \item Less than once a month 
        \item Never
    \end{enumerate}
    
    \item Second most accessed social network:
    \item Frequency of access:
    \begin{enumerate}
        \item Several times per day 
        \item Once a day
        \item Several times per week 
        \item Several times per month 
        \item Less than once a month 
        \item Never
    \end{enumerate}
    
    \item Third most accessed social network:
    \item Frequency of access:
    \begin{enumerate}
        \item Several times per day 
        \item Once a day
        \item Several times per week 
        \item Several times per month 
        \item Less than once a month 
        \item Never
    \end{enumerate}
    
\end{enumerate}

\subsubsection*{Posting your content}
In what social networks do you post your content (text, audio, videos, images, memes or GIFs) and how often do you post it?
(Provide information about the 3 social networks you post more often - these can be the same or different from the ones in the previous question)

\begin{enumerate}
  \setcounter{enumi}{24}
    \item First social network in which you most post your content:
    \item Frequency of posting:
    \begin{enumerate}
        \item Several times per day 
        \item Once a day
        \item Several times per week 
        \item Several times per month 
        \item Less than once a month 
        \item Never
    \end{enumerate}
    
    \item Second social network in which you most post your content:
    \item Frequency of posting:
    \begin{enumerate}
        \item Several times per day 
        \item Once a day
        \item Several times per week 
        \item Several times per month 
        \item Less than once a month 
        \item Never
    \end{enumerate}
    
    \item Third social network in which you most post your content:
    \item Frequency of posting:
    \begin{enumerate}
        \item Several times per day 
        \item Once a day
        \item Several times per week 
        \item Several times per month 
        \item Less than once a month 
        \item Never
    \end{enumerate}
    
\end{enumerate}

\subsubsection*{Sharing content from others}
In what social networks do you share content from others (text, audio, videos, images, memes or GIFs) that you authored and how often do you share it?
(Provide information about the 3 social networks where you share with more often - these can be the same or different from the ones in the previous questions)

\begin{enumerate}
  \setcounter{enumi}{30}
    \item First social network in which you most share content from others:
    \item Frequency of sharing:
    \begin{enumerate}
        \item Several times per day 
        \item Once a day
        \item Several times per week 
        \item Several times per month 
        \item Less than once a month 
        \item Never
    \end{enumerate}
    
    \item Second social network in which you most share content from others:
    \item Frequency of sharing:
    \begin{enumerate}
        \item Several times per day 
        \item Once a day
        \item Several times per week 
        \item Several times per month 
        \item Less than once a month 
        \item Never
    \end{enumerate}
    
    \item Third social network in which you most share content from others:
    \item Frequency of sharing:
    \begin{enumerate}
        \item Several times per day 
        \item Once a day
        \item Several times per week 
        \item Several times per month 
        \item Less than once a month 
        \item Never
    \end{enumerate}    
    
\end{enumerate}

\subsubsection*{Posting your content}
Please tell us how often you post your content on social networks, according to the following types of content:

\begin{enumerate}
  \setcounter{enumi}{36}
    \item Text: 
    \begin{enumerate}
        \item Several times per day 
        \item Once a day
        \item Several times per week 
        \item Several times per month 
        \item Less than once a month 
        \item Never
    \end{enumerate}
    
    \item Audio: 
    \begin{enumerate}
        \item Several times per day 
        \item Once a day
        \item Several times per week 
        \item Several times per month 
        \item Less than once a month 
        \item Never
    \end{enumerate}
    
    \item Video: 
    \begin{enumerate}
        \item Several times per day 
        \item Once a day
        \item Several times per week 
        \item Several times per month 
        \item Less than once a month 
        \item Never
    \end{enumerate}
    
    \item Image: 
    \begin{enumerate}
        \item Several times per day 
        \item Once a day
        \item Several times per week 
        \item Several times per month 
        \item Less than once a month 
        \item Never
    \end{enumerate}
    
    \item Meme: 
    \begin{enumerate}
        \item Several times per day 
        \item Once a day
        \item Several times per week 
        \item Several times per month 
        \item Less than once a month 
        \item Never
    \end{enumerate}
    
    \item GIF: 
    \begin{enumerate}
        \item Several times per day 
        \item Once a day
        \item Several times per week 
        \item Several times per month 
        \item Less than once a month 
        \item Never
    \end{enumerate}
    
\end{enumerate}

\subsubsection*{Sharing content from others}
Please tell us how often you share content from others on social networks, according to the following types of content:

\begin{enumerate}
  \setcounter{enumi}{42}
    \item Text: 
    \begin{enumerate}
        \item Several times per day 
        \item Once a day
        \item Several times per week 
        \item Several times per month 
        \item Less than once a month 
        \item Never
    \end{enumerate}
    
    \item Audio: 
    \begin{enumerate}
        \item Several times per day 
        \item Once a day
        \item Several times per week 
        \item Several times per month 
        \item Less than once a month 
        \item Never
    \end{enumerate}
    
    \item Video: 
    \begin{enumerate}
        \item Several times per day 
        \item Once a day
        \item Several times per week 
        \item Several times per month 
        \item Less than once a month 
        \item Never
    \end{enumerate}
    
    \item Image: 
    \begin{enumerate}
        \item Several times per day 
        \item Once a day
        \item Several times per week 
        \item Several times per month 
        \item Less than once a month 
        \item Never
    \end{enumerate}
    
    \item Meme: 
    \begin{enumerate}
        \item Several times per day 
        \item Once a day
        \item Several times per week 
        \item Several times per month 
        \item Less than once a month 
        \item Never
    \end{enumerate}
    
    \item GIF: 
    \begin{enumerate}
        \item Several times per day 
        \item Once a day
        \item Several times per week 
        \item Several times per month 
        \item Less than once a month 
        \item Never
    \end{enumerate}
    
\end{enumerate}

\subsection*{Social networks accessibility practices}
Several social networks currently offer their users the possibility of providing alternative text descriptions for visual elements in order to improve the understanding of these contents by people with visual disabilities.

\begin{enumerate}
  \setcounter{enumi}{48}
    \item For this question consider the last three posts you have authored with any kind of media content (i.e.: audio, videos, photos, memes and GIFs).
    Did you provide an alternative text description for any of these media contents? 
    \begin{enumerate}
        \item Yes
        \item No
    \end{enumerate}
    
    \item Why didn't you provide an alternative text description for any of these media contents? (Choose all that apply)
    \begin{enumerate}
        \item I didn't knew it was possible
        \item I don't know how to write a suitable alternative text description 
        \item I don’t know where to write an alternative text description
        \item It’s too time consuming
        \item Other:
    \end{enumerate}
    
    \item In which social network? (Choose all that apply)
    \begin{enumerate}
        \item Facebook
        \item Twitter
        \item Instagram
        \item Other:
    \end{enumerate}    
    
    \item How did you provide an alternative text description for any of these media contents? (Choose all that apply)
    \begin{enumerate}
        \item Writing the description in the text of the post
        \item Using the functionality provided by the social network    
        \item Using an automatic service for alternative text generation
        \item Other:
    \end{enumerate}    
        
    \item Why do you think social network users do not provide alternative text descriptions for the media content they post or share? (Choose all that apply)
    \begin{enumerate}
        \item They don't know it is possible
        \item They don't know how to write a suitable alternative text description 
        \item They don’t know where to write an alternative text description
        \item They consider it is too time consuming
        \item They don’t think it has any impact
        \item Other:
    \end{enumerate}   
    
    \item Besides improving access and understanding of the content by people with disabilities, what are or would be your motivations for posting accessible media content on social networks?

    \item When considering posting media content do you take any additional steps to enhance the content accessibility?
    \begin{enumerate}
        \item Yes
        \item No
    \end{enumerate}

    \item Which ones?
    
    \item In which social networks? (Choose all that apply)
    \begin{enumerate}
        \item Facebook
        \item Twitter
        \item Instagram
        \item Other: \underline{\hspace{3cm}}        
    \end{enumerate}
    
    \item Are you aware of if your family, friends or acquaintances have any concerns about accessibility when posting or sharing content in social media ?
    \begin{enumerate}
        \item Yes
        \item No
        \item I don't know
        \item Prefer not to answer
    \end{enumerate}    
    
    \item Which ones?
    
    \item Would you like to have access to a tool that would support you in the authoring of accessible content?
    \begin{enumerate}
        \item Yes
        \item No
    \end{enumerate}

    \item Which features would you include?
    
    \item Do you have some additional thoughts or comments about the subject that you would like to share?
    
    \item The next phase of this research will consist of a second study where participants will be invited to describe in greater detail the difficulties felt when posting any kind of digital media, taking into account the accessibility of this content.
    If you are interested in participating in this next stage of the study, please provide us with your email address so we can provide you with further details:
    
\end{enumerate}    

\end{document}